\documentclass[10pt,english,aps,prb,twocolumn]{revtex4-2}
\usepackage[T1]{fontenc}
\usepackage[latin9]{inputenc}
\usepackage{float}
\usepackage{dsfont}
\usepackage{bm}
\usepackage{amssymb}
\usepackage{graphicx}
\usepackage{xcolor}

\makeatletter

\providecommand{\tabularnewline}{\\}

\makeatother

\usepackage{babel}
\begin{document}
\title{Strain Engineering of Photo-induced Topological Phases in 2D Ferromagnets}
\author{T. V. C. Antão$^{1}$ and N. M. R. Peres$^{2,3,4}$}
\affiliation{$^{1}$Laboratório de Instrumentação e Física Experimental de Partícuals (LIP), University of Minho, 4710-057 Braga, Portugal}
\affiliation{$^{2}$Centro de Física das Universidades do Minho e do Porto (CF-UM-UP) e Departamento de Física, Universidade do Minho, P-4710-057 Braga,
Portugal}
\affiliation{$^{3}$International Iberian Nanotechnology Laboratory (INL), Av Mestre
José Veiga, 4715-330 Braga, Portugal}
\affiliation{$^{4}$POLIMA---Center for Polariton-driven Light-Matter Interactions, University of Southern Denmark, Campusvej 55, DK-5230 Odense M, Denmark}
\begin{abstract}
We argue that strain engineering is a powerful tool which may facilitate
the experimental realization and control of topological phases in
laser-driven 2D ferromagnetic systems. To this extent, we show that
by applying a circularly polarized laser field to a 2D honeycomb ferromagnet
which is uniaxially strained in either the zig-zag or armchair direction,
it is possible to generate a synthetic Dzyaloshinskii-Moriya interaction
(DMI) tunable by the intensity of the applied electric field, as well
as by the magnitude of applied strain. Such deformations enable transitions
to phases with opposite sign of Chern number, or to trivial phases.
These are basic results that could pave the way for the development
of a new field of Strain Engineered Topological Spintronics (SETS). 
\end{abstract}
\maketitle

\section{Introduction}

With the experimental observation of magnetic order in two-dimensional
(2D) materials in 2017 \cite{gong_discovery_2017,huang_layer-dependent_2017}
and the simultaneous growth in interest in topological aspects of
condensed matter systems over the past decade, the ability to generate,
study, and manipulate topological phases of magnetic materials has
become a rapidly growing research direction. Systems such as Magnon
Chern Insulators or other varieties of topological magnetic systems
have been theoretically studied \cite{zhang_topological_2013} and
experimentally verified in the past few years \cite{cai_topological_2021,chisnell_topological_2015}.
The reason for this interest is that topology describes effects which
stem from global properties of the band structure robust to small
local perturbations, such as impurities, and can have a profound effect
on a the material physical properties. Topological insulating phases
are characterized, for instance, by the existence of chiral edge states
with high mobility. Their robustness is desirable for a variety of
applications such as spintronics, and ensuing technological implementations
\cite{he_topological_2022}.

It is evident that for such applications, the ability to manipulate
topological phases, whether by switching topological properties on
or off, or alternating between distinct topological phases, are desirable
goals.  

{Additionally, magnon based approaches to spintronic technologies have also gained traction for a variety of reasons \cite{Zhixiong_2023_Topology_in_Collective_Magnetization_Dynamics}: From their ability to propagate without generating electrical current and therefore reducing losses, to the possibility of making use of their internal degrees of freedom to implement logic gates \cite{Schneider_2008_Spin_Wave_Logic_Gates, Balynskiy_2018_Magnetic_Logic_Gates}, and to their large diffusion lengths in comparison to electrons \cite{Cornelissen_2015_long_distance1, Pirro_2014_Spin_Wave_Propagation,liu_2018_long-distance}, magnons have garnered attention as a convenient excitation for processing and transporting information. For this reason, magnon spintronics relies on the use of magnons as intermediate agents, being that information initially coded in charge or spin of electrons can be converted to magnon currents, subsequently dispatched to and handled at potentially different devices, and finally converted back. In combination with the attractiveness of topology, the use of magnons renders the study of topological spin systems a worthwhile endeavor for the development of spintronic devices \cite{Wang_2017_Topologically_protected,Wang_2018_Topological_Magnonics}.}

One possibility for engineering topology in spin systems relies
on the fact that a 2D ferromagnet with a honeycomb lattice structure
which hosts a strong intrinsic Dzyaloshinskii-Moriya interaction (DMI)
\cite{dzyaloshinsky_thermodynamic_1958,moriya_anisotropic_1960}
can have the magnitude of this interaction renormalized when irradiated
by a circularly polarized laser field \cite{owerre_floquet_2017,owerre_corrigendum:_2018}.
Indeed, It was been predicted that CrI$_{3}$ hosts topological magnons
at the $\mathbf{K}-$point of the hexagonal Brillouin zone, starting
from a full electronic model \cite{Costa_2020}.

A Heisenberg spin model (HSM) with a DMI provides a concrete realization of the Haldane model \cite{haldane_model_1988} for magnonic excitations.
This model is known to host edge states, which are the hallmark of
Chern insulator phases. Electronic Chern insulators, for instance,
are able to conduct electrons along their edges and yet remain insulators
in their bulk, and in a similar manner, magnonic insulator samples
host gapless bands for spin excitations along their edges while remaining
gapped in the bulk. This behavior results in a measurable thermal Hall response \cite{katsura_theory_2010,Hosho_2010_Thermal_Hall,onose_observation_2010,matsumoto_theoretical_2011,matsumoto_thermal_2014,han_spin_2017}
. A field-dependent renormalization of the DMI can result in the possibility
of the topological properties of spin systems being manipulated, such
as the direction of edge state conduction being reversed or entirely
switched off, by changing the magnitude of the applied electric field.
In addition, a DMI resulting entirely from the interaction of laser
fields with spins \cite{owerre_floquet_2017} can also be generated.
In this manner, if a material does not naturally host such an interaction,
it can be synthesized by a laser beam, yielding a so-called Floquet
Magnon Chern insulator (FMCI). In case the material's intrinsic DMI
is weak compared to such a synthetic term, the control of topological
properties is limited: Increasing the intensity of the laser field
can turn the interaction on or off, {but doesn't provide a way to reverse edge spin states or additional desirable features}. Besides, this limited
tuning occurs only for very precise (and large) values of the intensity
of the applied fields. Thus, if it were possible to induce this interaction
in a fully tunable manner to a larger class of materials, one expects
that new technological developments based on topological spintronics
could arise. This paper addresses the manipulation of these topological
states by proposing a method based on elastically deforming, i.e.
straining, a 2D magnetic material. Strain can be applied in a variety
of ways, including the deposition of a 2D material onto, and subsequent
deformation of an elastic substrate \cite{peng_strain_2020}. It
has proven to be an extremely powerful tool in semiconductors, as
well as in 2D materials such as graphene, where band structure properties
can be manipulated \cite{low_gaps_2011,pereira_tight-binding_2009},
and other electronic properties can be locally changed using patterned
substrates. These patterns, such as bends or folds, wells, bubbles,
and troughs can induce mechanical strain on an overlaid mono-layer
of material, and may be used to design all-graphene integrated circuits
\cite{pereira_strain_2009-1}. In Cr$_{2}$Te$_{3}$, strain engineered
magnetism has been observed \cite{zhong_strain-sensitive_2022} and
when it comes to topology, strain in the Haldane model has also been
theoretically considered in the past \cite{mannai_strain_2020},
where it has been shown to be able to induce topological phase transitions
to a trivial state. The DMI can also be subject to changes due to
strain \cite{udalov_strain-dependent_2020}, and this conjugation
of factors is a good indicator that strain is a useful tool when considering
topological properties of magnetic materials. We show here that this
is indeed the case, as straining a 2D ferromagnet irradiated by a
laser field can invert the sign of its topological invariant, as well
as {induce a transition} from a topologically insulating phase to the trivial phase.

Our calculations, lying at the interface between Strain Engineering
and Floquet Engineering may pave the way for a new class of Strain
Engineered Topological Spintronic (SETS) devices, based on local applications
of strain to ferromagnetic 2D materials, as in this work, we propose
a mechanism for the realization of tunable photo-induced topology
in a large class of 2D ferromagnetic materials based on strain. We
draw phase diagrams based on the computation of the Floquet Chern
number for a FMCI in the honeycomb lattice, as a function of tensile
strain and the magnitude of an applied laser field which clearly exhibit
strain-driven transitions. We consider two main cases: First, a next-nearest
neighbor (NNN) interaction is given by a DMI alone; and second, an
extension of this model where one also considers an additional NNN
Heisenberg coupling. We show that strain-induced topological phase
transitions occur in both systems. However, due to the mapping between
the latter model and an anisotropic Haldane model with the key property
of tunable fluxes, phase transitions can occur for smaller electric
field intensities, and small amounts of strain. We argue that such
a model could provide the breeding ground for new developments based
on SETS.

\section{Unstrained model Hamiltonian and limitations}

\subsection{The Floquet Magnon Chern Insulator}

\begin{figure}
\includegraphics[scale=0.105]{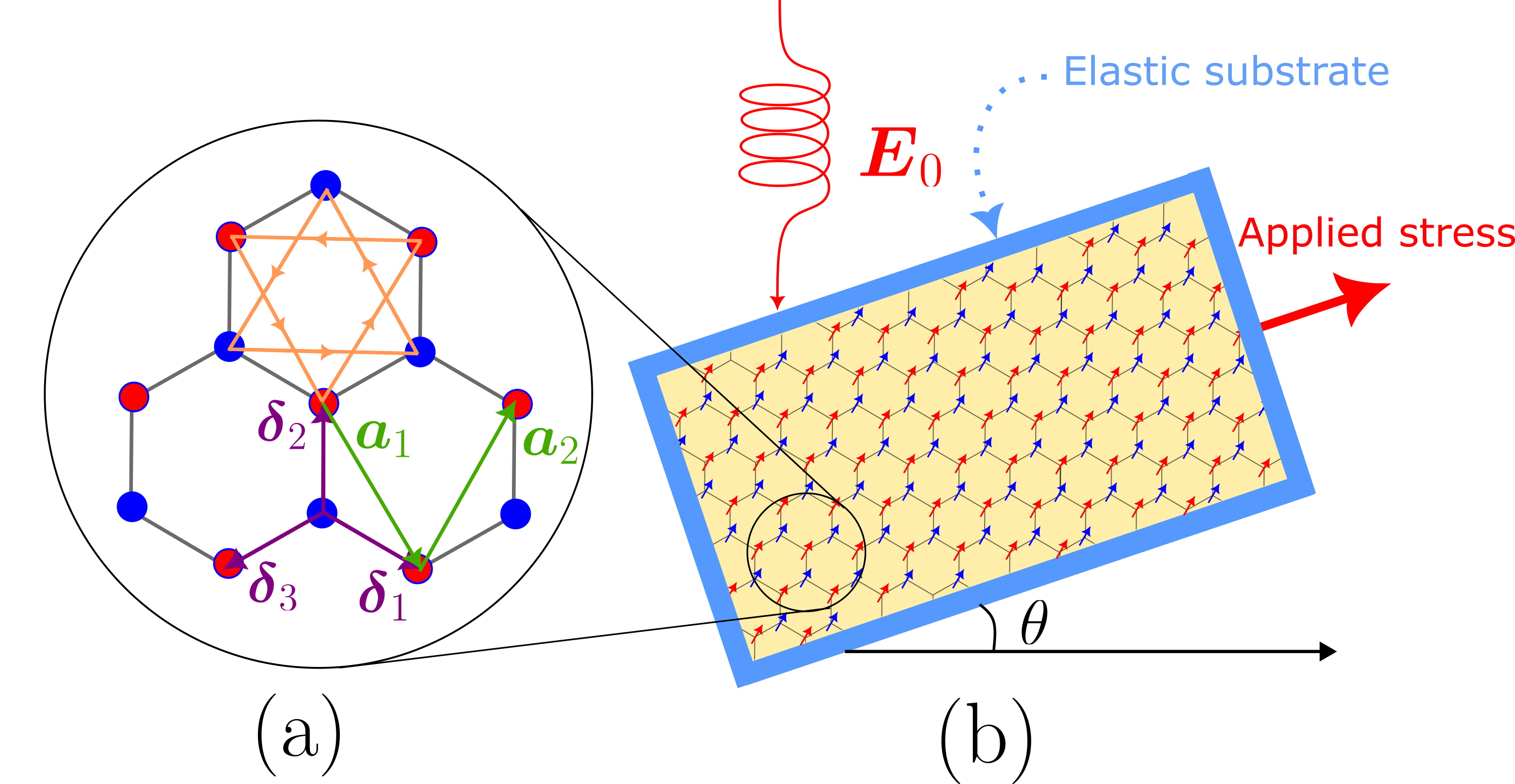}

\caption{(a) Honeycomb lattice structure: A (B) sub-lattice atoms are marked
in blue (red). NN vectors $\bm{\delta}_{i}$ are displayed
in purple and NNN vectors $\bm{a}_{i}$ are displayed in green.
Orange arrows in the topmost honeycomb shape showcase the flux factor
of the Dzyaloshinskii-Moriya interaction. (b) Schematic for the setup
for straining a ferromagnetic 2D material at an angle $\theta$ if
stress is applied along the large blue arrow. An external electric
circularly polarized electric field is applied with magnitude $E_{0}$
leading to the tunable topological properties described in the main
text. Throughout the text we will consider $\theta=0$ corresponding
to zig-zag (ZZ) strain, and $\theta=\pi/2$, corresponding to armchair
(AC) strain \label{fig:strain_steup}}
\end{figure}

We start by describing the basics of the magnetic model that serves
as the basis for our proposal by considering the previously discussed
FMCI in a honeycomb lattice of spin $S$ atoms. The structure of the
honeycomb lattice is given in panel (a) of Fig. \ref{fig:strain_steup}.
It is a Bravais lattice with two atoms per unit cell and thus can
be thought of as being composed of two distinct sub-lattices, which
we label by $A$ and $B$. An $A-$sub-lattice atom is connected to
its nearest neighbors (NN) via the vectors 
\begin{eqnarray}
\bm{\delta}_{1} & =&(\sqrt{3}/2,-1/2)a_{0},\\
\bm{\delta}_{2} & =&(0,1)a_{0},\\
\bm{\delta}_{3} & =&(-\sqrt{3}/2,-1/2)a_{0},
\end{eqnarray}
where $a_{0}$ is the inter-atomic distance, in a pristine, unstrained
lattice. A 2D honeycomb ferromagnet can be described by a HSM Hamiltonian
which depends on the spin vector operators $\bm{S}(\bm{r}_{i})=\left(S^{x}(\bm{r}_{i}),S^{y}(\bm{r}_{i}),S^{z}(\bm{r}_{i})\right)$,
acting at position $\bm{r}_{i}$ which can couple to spins
at NN and NNN sites via in-plane exchange integrals $J_{\perp}$ and
$J_{2,\perp}$, respectively, as well as their $J_{z}$ and $J_{z,2}$
counterparts. The Hamiltonian reads explicitly

\begin{eqnarray}
H&= & \frac{1}{2}\sum_{\left\langle i,j\right\rangle }J_{\perp}S^{+}(\bm{r}_{i})S^{-}(\bm{r}_{j})+\mathrm{h.c.}\nonumber\\
  &+&\frac{1}{2}\sum_{\langle\langle i,j\rangle\rangle}J_{2,\perp}S^{+}(\bm{r}_{i})S^{-}(\bm{r}_{j})+\mathrm{h.c}\nonumber\\
  &+&\sum_{\left\langle i,j\right\rangle }J_{z}S^{z}(\bm{r}_{i})S^{z}(\bm{r}_{j})\nonumber\\
  &+&\sum_{\langle\langle i,j\rangle\rangle}J_{2,z}S^{z}(\bm{r}_{i})S^{z}(\bm{r}_{j}),
\end{eqnarray}
where $\left\langle \cdot,\cdot\right\rangle $ indicate a restriction
of the summation to NN sites and $\langle\langle\cdot,\cdot\rangle\rangle$
a restriction to NNN sites. In addition, a representation in terms
of the spin ladder operators $S^{\pm}=S^{x}\pm iS^{y}$ is used. This
model hosts gapless Dirac magnons (quantized spin-waves) excitations
whose bands showcase an ultra-relativistic dispersion near the $\bm{K}$
and $\bm{K}'$ points of the Brillouin zone.

An FMCI is built from a HSM quite subtly, as the fact that an electric
field alone can couple to magnons, is, in principle, not so obvious.
The key effect that comes into play, which allows for the direct coupling
of an electric field to neutral bosons which carry a magnetic moment,
such as Dirac magnons, is the Aharonov-Casher (A-C) effect \cite{aharonov_topological_1984, ASTPires_2021_Theoretical_Spin}.
This is a dual effect to the more well known Aharonov-Bohm (A-B) effect
\cite{aharonov_significance_1959}, in which a charged particle in
a region of space with zero magnetic field, but importantly non-zero
magnetic vector potential, acquires a non-trivial topological phase.
Both A-B and A-C effects can result in interference, and the
A-C effect implies that a charge neutral particle with a magnetic
moment moving in a electric field will also acquire such a phase,
called the A-C phase. For a ferromagnet irradiated by a circularly
polarized laser field $\bm{E}(t)=E_{0}\left(\tau\cos\omega t,\sin\omega t,0\right)$,
with handedness given by $\tau=\pm1$, and with frequency $\omega$,
the A-C phase manifests itself as a time-dependent Peierls phase acquired
by the Dirac magnons when hopping between different lattice sites. 

The time dependence of the resulting Hamiltonian may appear initially
cumbersome, as well as not particularly elucidating as to the underlying
physical effects the Dirac magnons experience. For these reasons,
a perturbative scheme has historically been considered for the analysis
of such Hamiltonians, based on the analysis of periodically driven
systems. This is the so-called Floquet theory. Using the Floquet theory
framework, it is possible to perform a high-frequency expansion in
inverse powers of $\omega$ \cite{eckardt_high-frequency_2015} which
provides an effective Hamiltonian up to $\mathcal{O}(\omega^{-1})$ with clear qualitative and physical
interpretation. The correction of
lowest order $\mathcal{O}\left(\omega^{0}\right)$ in high-frequency
provides an averaging of the Hamiltonian over a period of the driving
laser, yielding a renormalization of the NN and NNN in-plane exchange
integrals as $J_{\perp}\to J_{\perp}\mathcal{J}_{0}(\tau\alpha a_{0}),J_{\perp,2}\to J_{\perp,2}\mathcal{J}_{0}(\tau\alpha\sqrt{3}a_{0})$,
(Eq. \ref{eq:Eq1}) where $\alpha\equiv\mu_{B}E_{0}/\hbar c^{2}$,
with $\mu_{B}$ the Bohr magneton. $\mathcal{J}_{n}(x)$ are the $n$th
order Bessel functions, and the constants $\hbar$ and $c$ are the
reduced Planck's constant and the speed of light in vacuum, respectively. The functional
form of the renormalization of the NN in-plane hoppings is already
interesting despite not providing topological properties by itself,
as it depends on a Bessel function of order 0 for the case of the
NN hoppings. This allows for tuning between Heisenberg type-couplings
and Ising type couplings, for instance, since all $J_{\perp}$ can
be turned off. In the literature one often defines the dimensionless
parameter $\lambda=\alpha a_{0}$, however, we make here explicit
the additional knob for this model, which is the key ingredient to
the results showcased in this work, is the inter-atomic distance $a_{0}$.

Besides this first order correction, corresponding to the renormalization
of $J_{\perp}$ and $J_{\perp,2}$, the second-order high-frequency
correction can be seen to yield an additional term in the effective
Hamiltonian, in the form of a spin-chirality (Eq. \ref{eq:Eq2}).
The Hamiltonian will read $H_{F}=\hat{H}_{F}^{(1)}+\hat{H}_{F}^{(2)}+\mathcal{O}\left(1/\omega^{2}\right)$,
and the two first terms in the high-frequency expansion explicitly
read

\begin{eqnarray}
\hat{H}_{F}^{(1)}&=& -\sum_{\left\langle i,j\right\rangle }\frac{J_{\perp}\mathcal{J}_{0}(\tau\alpha a_{0})}{2}S^{+}(\bm{r}_{i})S^{-}(\bm{r}_{j})+\mathrm{h.c}\nonumber \\
  &-&\sum_{\left\langle i,j\right\rangle }J_{z}S^{z}(\bm{r}_{i})S^{z}(\bm{r}_{j})-\sum_{\left\langle \left\langle i,j\right\rangle \right\rangle }J_{2,z}S^{z}(\bm{r}_{i})S^{z}(\bm{r}_{j})\nonumber \\
  &-&\sum_{\left\langle \left\langle i,j\right\rangle \right\rangle}\frac{J_{2,\perp}\mathcal{J}_{0}\left(\tau\alpha\sqrt{3}a_{0}\right)}{2}S^{+}(\bm{r}_{i})S^{-}(\bm{r}_{j})+\mathrm{h.c}\label{eq:Eq1}\\
\hat{H}_{F}^{(2)}&= & \sum_{\left\langle i,\langle j\rangle,k\right\rangle }\chi_{ijk}\bm{S}(\bm{r}_{i})\cdot\left(\bm{S}(\bm{r}_{j})\times\bm{S}(\bm{r}_{k})\right).\label{eq:Eq2}
\end{eqnarray}
Here, the interlinked braces $\left\langle \cdot,\langle\cdot\rangle,\cdot\right\rangle $
indicate that the summation is performed over NNN atoms at positions
$i$ and $k$, which are connected by position $j$. In Eq. \ref{eq:Eq2},
the spin-chirality is given in magnitude by 
\begin{equation}
\chi_{ijk}=\tau\sqrt{3}J^{2}\mathcal{J}_{1}(\tau\alpha\delta_{ji})\mathcal{J}_{1}(\tau\alpha\delta_{ik})\nu_{ik}^{A/B}/\omega\,,
\end{equation}
where $\nu_{ik}^{A}=-\nu_{ik}^{B}=-\nu_{ki}^{A}$ is a flux-like term,
dependent on the orientation of the NNN bonds which connect sites
$i$ and $k$ according to the orange arrows in Fig. \ref{fig:strain_steup},
panel (a). Note that we have made clear the fact that the intensity
of the spin-chirality depends on the successive hoppings between an
intermediate site via the first order Bessel functions. At this stage,
all the distances $\delta_{ji}\equiv|\bm{\delta}_{i}|a_{0}=a_{0}$
are the same, and equal the inter-atomic distance. As such, we can
write $\mathcal{J}_{1}(\tau\alpha\delta_{ji})\mathcal{J}_{1}(\tau\alpha\delta_{ik})=\mathcal{J}_{1}^{2}(\tau\alpha a_{0})$.
This makes it so $\mathrm{sign}\left(\chi\right)=\tau$, for any possible
value of $\alpha\propto E_{0}$. Such a spin chirality term is known
to originate frustration in the ground state of the ferromagnetic
system, leading to the possibility of originating spin-liquid states \cite{Wen_Wilczek_Zee_1989_Chiral_Spin_Liquids},
but considering $J_{z}>J_{\perp}$ the ferromagnetic ground state
is stabilized.

For pursuing our discussion, a second quantization formalism for magnons
can be employed using the Holstein-Primakoff (HP) bosonization. The
(linearized) HP transformations map spin operators into bosonic creation
and annihilation operators $a_{i}^{\dagger}/a_{i}$ ($b_{i}^{\dagger}/b_{i}$)
within the A (B) sub-lattices. Within linear spin-wave theory, the
spin-chirality is indistinguishable from a DMI, and, indeed, when
writing this term using HP operators, and retaining only terms up
to second order in such operators, this equivalence becomes clear.
A particularly useful way to write this Hamiltonian in momentum space
can be achieved in terms of the Pauli matrices. If we consider the
vector of creation operators $\Psi_{\bm{k}}^{\dagger}=\left(a_{\bm{k}}^{\dagger},b_{\bm{k}}^{\dagger}\right)$,
the effective Hamiltonian in momentum space can be written as 
\begin{equation}
\mathcal{H}_{F}=\sum_{\bm{k}}\Psi_{\bm{k}}^{\dagger}\left[h_{0}(\bm{k})\mathds{1}+\bm{h}(\bm{k})\cdot\bm{\sigma}\right]\Psi_{\bm{k}},\label{eq:Dirac_Mat}
\end{equation}
where $\mathds{1}$ is the $2\times2$ identity matrix, and we have
defined a scalar $h_{0}(\bm{k})$ and a vector $\bm{h}(\bm{k})=\left(h_{x}(\bm{k}),h_{y}(\bm{k}),h_{z}(\bm{k})\right)$
which is contracted with the vector of Pauli matrices $\bm{\sigma}=\left(\sigma_{x},\sigma_{y},\sigma_{z}\right)$.
We will give an explicit form of the $\bm{h}(\bm{k})$
vector and the $h_{0}(\bm{k})$ scalar in a bit, however,
let us first draw a useful comparison to the well known Haldane model.
For now, we note that this is a general way to write a $2\times2$
Hermitian operator, which is useful for our purposes, since the vector
$\bm{h}(\bm{k})$ contains all the information necessary
for the topological characterization of the system. For now, we note
that it includes summations over the NN and NNN vectors, and thus,
it is expected that changing these vectors can have an effect on the
spectrum as well as on the topological properties of this model. In
the absence of a DMI, we have $h_{z}(\bm{k})=0$, the system
is gapless and hence in a trivial topological phase. Turning on the
DMI, the spectrum of Dirac magnons becomes gapped, and thus, this
interaction can be interpreted as providing Dirac magnons with a mass.
This system then falls into the category of Chern insulators, for
which the signature of topology is the Chern number or TKNN invariant
\cite{PhysRevLett.49.405}, which takes non-zero integer values if
the material is in a topological phase, and is zero for a trivial
phase. In case the topology is photo-induced, one calls it the Floquet
Chern number $C_{\eta}^{F}$, and in any case, it can be computed
as an integral over the Berry curvature $\Omega_{\eta}^{F}(\bm{k})=\frac{\eta}{2}\hat{\bm{h}}(\bm{k})\cdot\left(\partial_{k_{x}}\hat{\bm{h}}(\bm{k})\times\partial_{k_{y}}\hat{\bm{h}}(\bm{k})\right)$
in the full Brillouin zone, i.e.

\begin{equation}
C_{\eta}^{F}=\frac{1}{2\pi}\int_{\mathrm{BZ}}d^{2}\bm{k}\Omega_{\eta}^{F}(\bm{k}),
\end{equation}
In the expression for the Berry curvature and Floquet Chern number,
$\hat{\bm{h}}(\bm{k})=\bm{h}(\bm{k})/|\bm{h}(\bm{k})|$
and $\eta=\pm1$ is the band index. The Berry curvature itself can
also be computed from the eigenstates of the effective Hamiltonian
using numerical approaches \cite{fukui_chern_2005}, but the analytical
expression given above justifies the previous statement that the vector
$\bm{h}(\bm{k})$ contains all relevant information
necessary for the characterization of the material's topological properties.
For the unstrained FMCI, we have the analytical result $C_{\eta}^{F}=\eta\tau\mathrm{sign}\left(\chi\right)$,
using the explicit form of $\bm{h}(\bm{k})$ given
in subsection \ref{subsec:Mapping-the-FMCI}. Our numerical calculations
will employ Fukui's method \cite{fukui_chern_2005} due to its efficiency,
but they remain analytically verifiable. 

Finally, we can make our introductory comments about the manipulation
of topology being restricted in this model more precise. From the
analytical results for $C_{\eta}^{F}$ given a certain polarization
of light $\tau$, we have, so long as we set $J_{2,\perp}=0$, the
result $C_{\eta}^{F}=\eta\tau$, regardless of the intensity of the
laser field, with the exception of very special points, at which $\mathcal{J}_{1}^{2}(\tau\alpha a_{0})=0$,
at which $C_{\eta}^{F}=0$. The guiding motivation for the following
discussion is that this term stems from a more generic $\mathcal{J}_{1}(\tau\alpha\delta_{ji})\mathcal{J}_{1}(\tau\alpha\delta_{ik})$,
and hence, if inter-atomic distances could be changed, one could potentially
switch the sign of the Chern number. 

\subsection{Mapping the FMCI to a Bosonic Haldane model\label{subsec:Mapping-the-FMCI}}

In this section, we make explicit that the structure yielded by the
FMCI corresponds to bosonic Haldane model. This model is entirely
analogous to its fermionic counterpart with the exception of being
expressed at the cost {of} bosonic creation/annihilation operators in the
A and B sub-lattices of a honeycomb structure. Bosons are created
(annihilated) by $a_{i}^{\dagger}$($a_{i}$) in the A-sub-lattice
and by $b_{i}^{\dagger}$($b_{i}$) in the B-sub-lattice. It reads

\begin{eqnarray}
H&= & \sum_{i}M\left(a_{i}^{\dagger}a_{i}-b_{i}^{\dagger}b_{i}\right)+\sum_{\left\langle i,j\right\rangle }t_{ij}a_{i}^{\dagger}b_{i}+\mathrm{h.c.}\nonumber \\
 & +&\sum_{\left\langle \left\langle i,j\right\rangle \right\rangle }t_{2,ij}\left(a_{i}^{\dagger}a_{j}+b_{i}^{\dagger}b_{j}\right)+\mathrm{h.c.}\nonumber \\
 & +&i\sum_{\left\langle \left\langle i,j\right\rangle \right\rangle }t_{2,ij}'\left(a_{i}^{\dagger}a_{j}-b_{i}^{\dagger}b_{j}\right)+\mathrm{h.c.}
\end{eqnarray}
Here, $M$ is called the Haldane mass, and both real ($t_{2,ij})$
and imaginary $(i t_{2,ij}')$ NNN hoppings are present. One can bring
the two NNN hopping terms together by writing $t_{2,ij}+it_{2,ij}'=t_{2,ij}''e^{i\nu_{ij}\phi_{ij}},$where
$\phi_{ij}=\arctan\left(t_{2,ij}'/t_{2,ij}\right)$, and where $t_{2,ij}''=\sqrt{t_{2,ij}^{2}+t_{2,ij}'^{2}}$.
The factor $\nu_{ij}=\pm1$ is then chosen according to the direction
of the NNN hopping (see the orange arrows in Fig. 1). Thus, the Haldane
model Hamiltonian can be written, up to an arbitrary energy shift,
as

\begin{eqnarray}
H&=  &\sum_{i}M\left(a_{i}^{\dagger}a_{i}-b_{i}^{\dagger}b_{i}\right)+\sum_{\left\langle i,j\right\rangle }t_{ij}a_{i}^{\dagger}b_{i}+\mathrm{h.c.}\nonumber\\
  &+&\sum_{\left\langle \left\langle i,j\right\rangle \right\rangle }t_{2,ij}''e^{i\nu_{ij}\phi_{ij}}\left(a_{i}^{\dagger}a_{j}-b_{i}^{\dagger}b_{j}\right).
\end{eqnarray}
By writing the FMCI Hamiltonian, and specifically the second order
correction in terms of spin ladder operators, and subsequently using
the HP transformations, the spin-chirality term becomes a purely imaginary
NNN hopping. The identifications given in Table \ref{tab:Identifications-between-Haldane}
can then be performed. {Essentially, NN and NNN hoppings are mapped to their exchange integral counterparts, and the spin-chirality is mapped to an imaginary hopping, and a Haldane mass can exist for instance, in ferrimagnetic systems (where it is not actually $J_z$ which is different for the sublattices, but rather the value of the spin $S$) \cite{ASTPires_2021_Theoretical_Spin}.}
\begin{center}
\begin{table}[H]
\begin{centering}
\begin{tabular}{|c|c|}
\hline 
Haldane & FMCI\tabularnewline
\hline 
\hline 
$t$ & 3$J_{\perp}S\mathcal{J}_{0}(\alpha\delta_{ij})$\tabularnewline
\hline 
$t_{2}$ & $3J_{2,\perp}S\mathcal{J}_{0}(\alpha a_{ij})$\tabularnewline
\hline 
$t_{2}'$ & 6$J_{\perp}^{2}S^{2}\tau\mathcal{J}_{1}(\alpha a_{0})\mathcal{J}_{1}(\alpha a_{0})/\hbar\omega$\tabularnewline
\hline 
$M$ & $(J_{z,A}-J_{z,B})S/2$\tabularnewline
\hline 
\end{tabular}
\par\end{centering}
\caption{Identifications between Haldane model and Floquet Magnon Chern insulator{, for the case where no strain is present}.\label{tab:Identifications-between-Haldane}}
\end{table}
\par\end{center}

Furthermore, the flux $\phi_{ij}$ becomes dependent on the intermediary
site $k$ between $i$ and $j$, and reads $\phi_{ijk}=\arctan\left(\frac{J_{\perp}^{2}\tau}{J_{2,\perp}\hbar\omega}\frac{\mathcal{J}_{1}(\alpha a_{0})\mathcal{J}_{1}(\alpha a_{0})}{\mathcal{J}_{0}(\alpha\sqrt{3}a_{0})}\right),$ and
with this mapping underway, a momentum space representation can be
readily constructed by considering $\bm{h}(\bm{k})$
and $h_{0}(\bm{k})$ given by

\begin{eqnarray}
h_{0}(\bm{k}) & =&-\sum_{j}t_{2,j}''\cos\phi_{j}\cos\left(\bm{k}\cdot\bm{a}_{j}\right),\\
h_{x}(\bm{k}) & =&-\sum_{j}t_{j}\cos\left(\bm{k}\cdot\bm{\delta}_{j}\right),\\
h_{y}(\bm{k}) & =&\sum_{j}t_{j}\sin\left(\bm{k}\cdot\bm{\delta}_{j}\right),\\
h_{z}(\bm{k}) & =&M-2\sum_{j}t_{2,j}''\sin\phi_{j}\sin\left(\bm{k}\cdot\bm{a}_{j}\right).
\end{eqnarray}

{The inclusion of strain must now account for several physical phenomena, namely, it must describe changes in bond lengths and subsequent anisotropic variations in hopping and exchange interactions. In the following section a simple model which describes such variations is introduced.}

\section{Effects of strain on topological properties}

\subsection{FMCI with NN hoppings}

As we have previously described, for a magnetic material with a strong
intrinsic DMI the $\mathrm{sign}\left(\chi\right)$ can be manipulated
by changing the intensity of the laser field, due to the first order
correction of the high frequency approximation, which reads $\chi\to\mathcal{J}_{0}(\alpha a_{0})\chi$.
For systems where the DMI is fully synthetic and results only from
second order Floquet theory, this does not appear possible due to
the dependence in $\mathcal{J}_{1}^{2}(\tau\alpha a_{0})>0$. Guided
by the fact that a sign change can be achieved by making the system
anisotropic and transforming $\mathcal{J}_{1}^{2}(\tau\alpha a_{0})\to\mathcal{J}_{1}(\tau\alpha\delta_{ji})\mathcal{J}_{1}(\tau\alpha\delta_{ik})$,
we now explore the uniaxial straining of an FMCI, and show that even
small amounts of strain can provide a pathway for topological manipulations
of the model. Note that studies on the exchange parameters in CrI$_{3}$
as function of strain have been considered {and} theoretically analyzed
in the past, considering strain values up to $10\%$ \cite{Leon_2020}. 

\begin{figure*}
\begin{centering}
\includegraphics[scale=0.6]{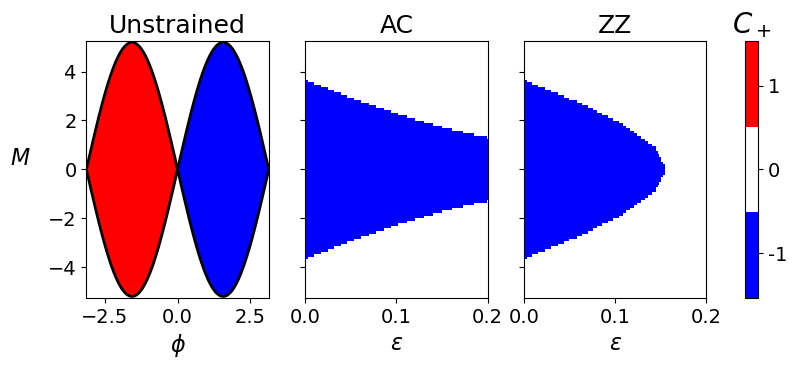}
\par\end{centering}
\caption{The leftmost panel showcases the phase diagram for a bosonic Haldane
model as a function of the Haldane mass $M$ and the phase $\phi$.
Black lines in this panel indicate $M=\pm3\sqrt{3}\sin\phi$, which
are the analytical results for the phase transition {lines} in this model.
The right panels showcase phase diagrams for the strained bosonic
Haldane model in the Zig-zag (ZZ) and Armchair (AC) directions. We
pick {$\beta=6.74$}, $\nu=0.165$ and a phase of $\phi=4\pi/5$, such
that for strain applied in the ZZ direction, for any value of $M$
and a strain above $\sim15$\%, the model is in a trivial phase. Generically,
for strain in the AC direction, no phase transitions are observed
for $M=0$. \label{fig:(a)-Phase-diagram}}
\end{figure*}

In the honeycomb lattice, the strain tensor is described by two parameters
alone, namely the tensile strain $\varepsilon$ and the Poisson ratio
$\nu$ \cite{landau_theory_2009}. When inducing stress onto the
2D magnetic material, the tensile strain $\varepsilon$ is proportional
to this stress, and therefore we can treat $\varepsilon$ as the tunable
parameter in our system. It measures the amount of deformation in
the direction of the applied stress, while the Poisson ratio measures
the deformation of the lattice in the transverse direction. A positive
Poisson ratio $\nu>0$ indicates that when a material is stretched
in a particular direction, it compresses in the transverse direction,
and vice-versa. As such, the lattice vectors acquire a functional
dependence on the parameters $\varepsilon$ and $\nu$, as strain
applied in a particular direction. The deformed vectors read $\bm{\delta}_{i}(\varepsilon,\nu,\theta)=\left(1+\overline{\overline{\varepsilon}}\right)\bm{\delta}_{i}^{(0)},$where
the strain tensor is

\begin{equation}
\overline{\overline{\varepsilon}}=
\left[
\begin{array}{cc}
\varepsilon\cos^{2}\theta-\nu\sin^{2}\theta & (1+\nu)\cos\theta\sin\theta\\
(1+\nu)\cos\theta\sin\theta & \sin^{2}\theta-\nu\cos^{2}\theta
\end{array}
\right].
\end{equation}
In the simplest tight-binding approach, strain can be included in
a given Hamiltonian via modifying hopping amplitudes anisotropically.
In previous works in the honeycomb lattice, it is considered that
electronic hoppings are exponentially suppressed when the bond length
is increased \cite{turchi_tight-binding_1998}. This is the simplest
possible model, which can be intuited phenomenologically from the
overlap of atomic orbitals. One has

\begin{eqnarray}
&&\mathrm{(NN):}\,  t_{ij}=t^{(0)}e^{-\beta\left(\delta_{ij}(\varepsilon,\nu,\theta)-1\right)},\label{eq:Eq1-1}\\
&&\mathrm{(NNN ):}\, t_{2,ij}=t_{2}^{(0)}e^{-\beta\left(a_{ij}(\varepsilon,\nu,\theta)-\sqrt{3}\right)},\label{eq:Eq2-1}
\end{eqnarray}
where $\beta$ is a phenomenological parameter of the order of unity.
Since we expect that the strength of the exchange interaction is $J\propto t^{2}/U$
for $U$ representing the strength of on-site Coulomb repulsion in
some underlying electronic model, we consider that a similar exponential
decay occurs for $J$ with the rate of $2\beta$. We find that the
inclusion of such a phenomenological correction in a standard Haldane
model is enough to produce topological phase transitions, when a system
is strained in the Zig-zag direction with values of, for instance,
$\varepsilon\sim15\%$ for a Haldane flux of $\phi=4\pi/5$, due to
a fusion of the magnonic Dirac points. This is presented in Fig. \ref{fig:(a)-Phase-diagram}.

For a standard DMI system, if $J_{2,z/\perp}=0$, i.e. an Haldane
flux is present corresponding to $\phi=\pi/2$, the critical strain
necessary for a topological phase transition can be much higher, making
it unfeasible for realistic applications. On the other hand, rich
phase diagrams emerging from uniaxially straining a FMCI can appear.
This is due to the intricate dependence of NN and NNN hoppings on
Bessel functions, as presented in Table \ref{tab:Identifications-between-Haldane-1}.
We focus first on the simplest case, with $J_{2,z}=J_{2,\perp}=0$, {but will later show that the inclusion of these terms yields several advantages}.
\begin{center}
\begin{table}[H]
\begin{centering}
\begin{tabular}{|c|c|}
\hline 
Haldane & Strained FMCI\tabularnewline
\hline 
\hline 
$t$ & 3$J_{\perp}S\mathcal{J}_{0}(\alpha\delta_{ij})e^{-2\beta\left(\delta_{ij}-1\right)}$\tabularnewline
\hline 
$t_{2}$ & $3J_{2,\perp}S\mathcal{J}_{0}(\alpha a_{ij})e^{-2\beta\left(a_{ij}-\sqrt{3}\right)}$\tabularnewline
\hline 
$t_{2}'$ & 6$J_{\perp}^{2}S^{2}\tau\mathcal{J}_{1}(\alpha\delta_{ik})\mathcal{J}_{1}(\alpha\delta_{kj})e^{-2\beta\left(\delta_{ik}+\delta_{kj}-2\right)}/\hbar\omega$\tabularnewline
\hline 
$M$ & $(J_{z,A}-J_{z,B})S/2$\tabularnewline
\hline 
\end{tabular}
\par\end{centering}
\caption{Identifications between Haldane model and Strained Floquet Magnon
Chern insulator.\label{tab:Identifications-between-Haldane-1}}
\end{table}
\par\end{center}

\begin{figure}
\begin{centering}
\includegraphics[scale=0.4]{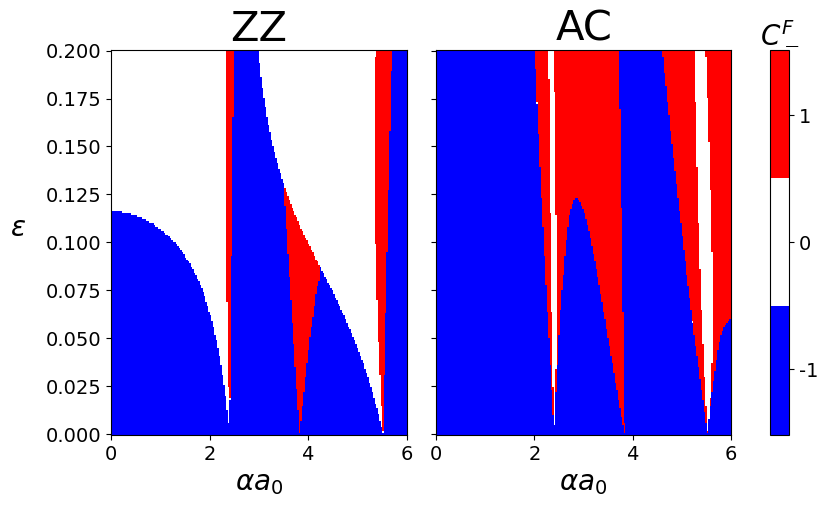}
\includegraphics[scale=0.18]{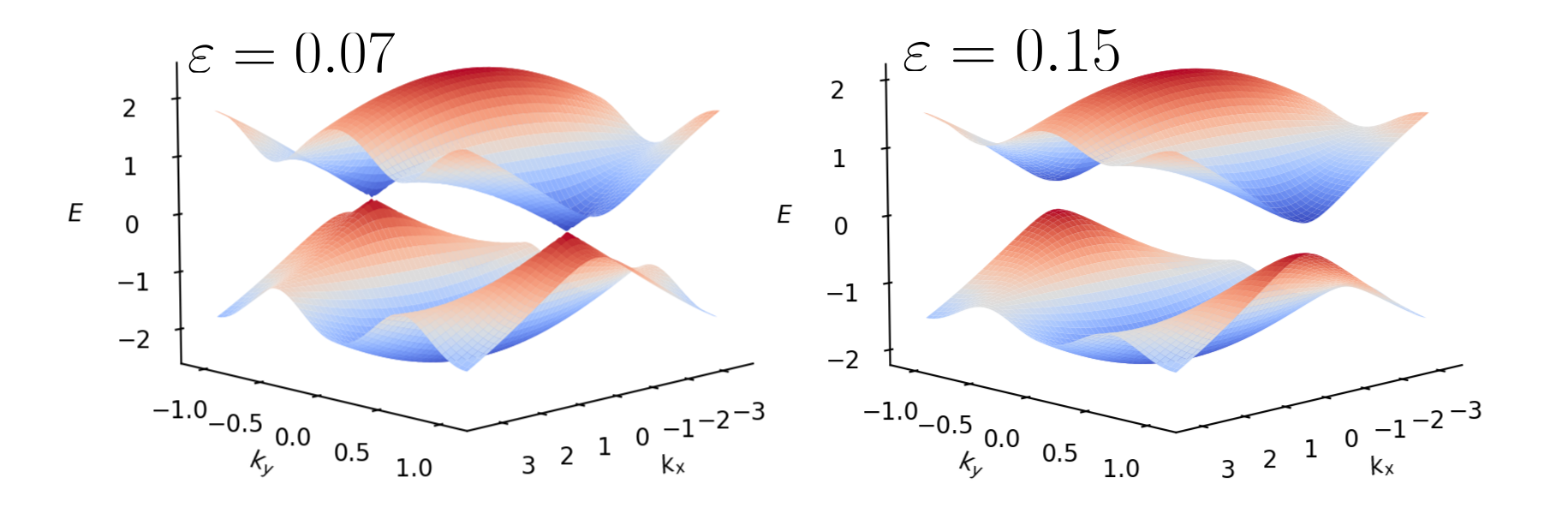}
\par\end{centering}
\caption{{Upper panels:} Phase diagrams showcasing the Chern number as a function of the electric
field through $\alpha a_{0}$ as well as the strain magnitude $\varepsilon$
in the ZZ and AC directions for a FMCI with $J_{2}=0$ and $\hbar\omega=50J$.
Left panel - Strain in the ZZ direction: A series of dips is observable
for which at certain values of electric field, small amounts of strain
are necessary to induce a topological phase transition. The first
white dip corresponds to a situation where $\mathcal{J}_{0}(\alpha|\delta_{1}|a_{0})$
and $\mathcal{J}_{0}(\alpha|\delta_{3}|a_{0})$ go to zero. The subsequent
transitions in each dip occur either because of this or due to changes
in the DMI sign due to $\mathcal{J}_{1}$. Right Panel - Strain in
the AC direction: A similar situation occurs, with the first dip being
related to the position of the zero of the NN exchange integrals,
and the second dips occur due to the DMI sign change. {Lower Panels: Band structure of the FMCI for a small value of electric field $\alpha a_0 = 0.01$ and strain along the ZZ direction. The left and right panels provide an example of the band structure below and above the critical strain, respectively. The transition from topological to trivial phases occurs due to the merging of the Dirac cones at the edge of the Brillouin zone. The topological gap in the left panel's band structure is practically invisible due to the small magnitude of electric field, which illustrates a difficulty inherent to the implementation of realizing photo-induced topological magnons with NN interactions alone.} \label{fig:Phase-diagrams-showcasing}}
\end{figure}

The functional dependence on Bessel functions can result in the closing
of the gap of the system well below strain values of $15$\%, and
break the symmetry of the lattice in such a way that the system becomes
topologically trivial, or even switch the relative sign of NN and
DMI. Indeed, this NN sign-switch plays a more relevant role at a lower
value of the intensity of the electric field, for any reasonable value
of strain. Thus, from an experimental point of view, may be more easily
accessed. Fig. \ref{fig:Phase-diagrams-showcasing} shows that both
if stress is applied in either the Zig-zag as well as Armchair directions
of the honeycomb lattice, there exist several points, near the zero
of $\mathcal{J}_{0}(\alpha|\delta_{1}(\varepsilon,\nu,\theta)|a_{0})=\mathcal{J}_{0}(\alpha|\delta_{3}(\varepsilon,\nu,\theta)|a_{0})$,
at which a transition between a Floquet Chern number $C_{-}^{F}=-1$
to $C_{-}^{F}=1$ can occur, even for very small values of strain.
For strain applied in the Armchair (AC) direction, a large region
of Chern number $C_{-}^{F}=+1$ occurs for strain above $12.5\%$
and $\alpha a_{0}$ above the first zero of $\mathcal{J}_{0}$, whereas
for strain applied in the ZZ direction, transitions to a trivial phase
can occur for a much smaller field intensity. 

The generic behavior of the transitions, is that several dips in critical
strain occur within the phase diagram, close to zeros of the Bessel
functions $\mathcal{J}_{0}$ and $\mathcal{J}_{1}$. As described
in the section on realistic parameter values, $\alpha a_{0}=2.3$,
which corresponds to the location of the first dip for which transitions
occur for low values of strain, is still quite a large field intensity.
This leads to the question of whether there exist any mechanism which
can lower this critical value further. 

\subsection{FMCI with NNN hoppings}

\begin{figure}
\begin{centering}
\includegraphics[scale=0.4]{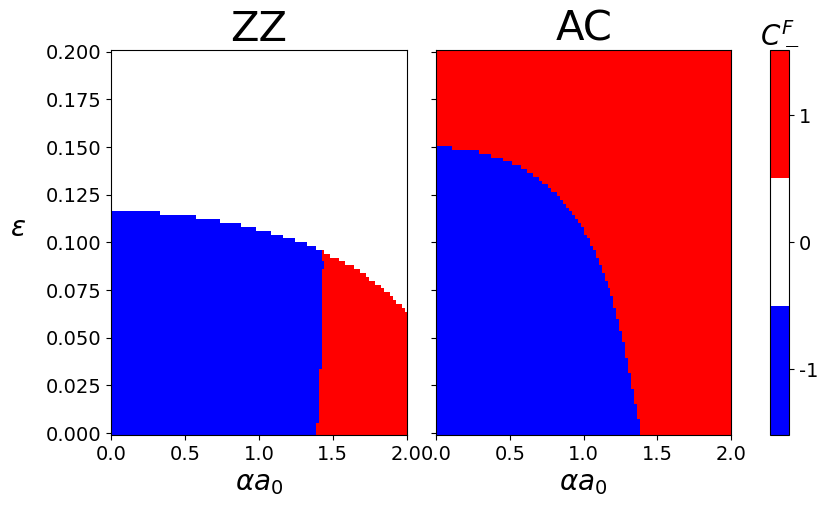}
\includegraphics[scale=0.18]{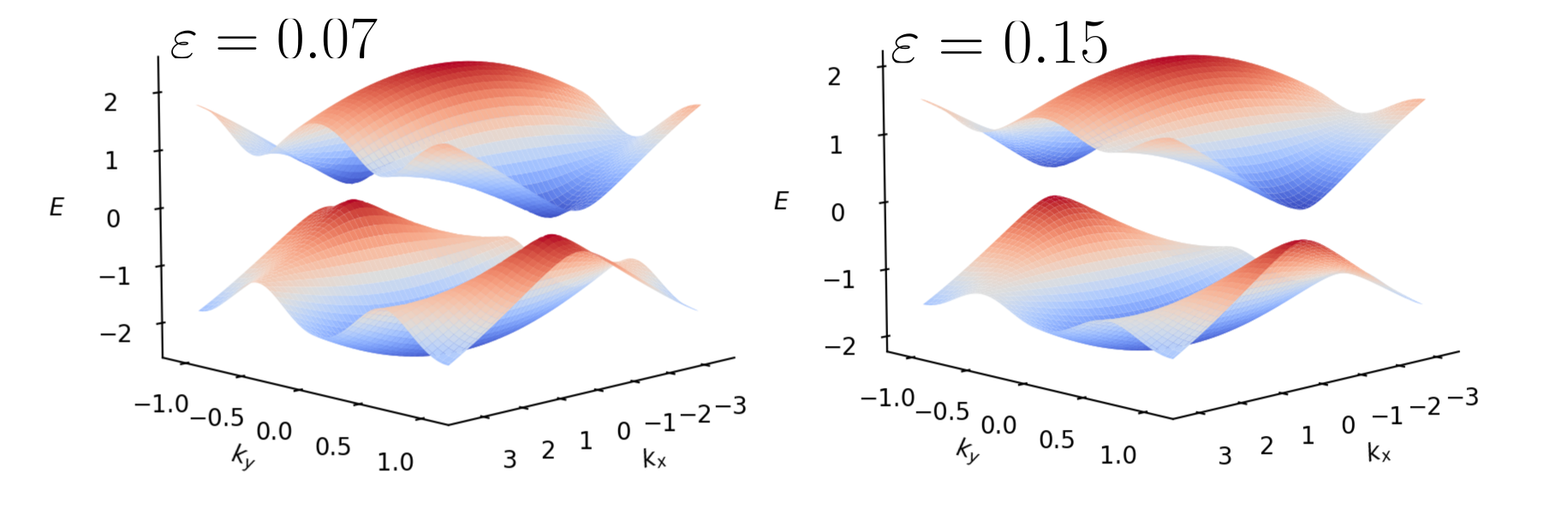}
\par\end{centering}
\caption{{Upper panels:} Phase diagrams showcasing the Chern number as a function of the electric
field through $\alpha a_{0}$ as well as the strain magnitude $\varepsilon$
in the Zig-zag (left panel) and Armchair (right panel) directions,
with $J_{2}=0.1J$ and $\hbar\omega=50J$. For strain in the Armchair
direction, there exist critical strain values for field arbitrarily
close to zero, as well as critical field values for arbitrarily small
strain near $\alpha a_{0}\approx1.38$, for which topological transitions
can occur between phases with inverse Chern numbers. {Lower panels: Band structure for a small value of electric field $\alpha a_0 = 0.01$ and strain in the ZZ direction. The left and right panels provide an example of the band structure below and above the critical strain, respectively. The presence of a NNN exchange integral $J_2>0$ clearly increases the gap (see Fig. \ref{fig:(a)-Phase-diagram}) and further stabilizes the topological phase, reducing the necessary values of electric fields for the realization of photo-induced topology.}\label{fig:Phase-diagrams-showcasing-1}}
\end{figure}

We now show that by turning on the NNN exchange integral $J_{2}>0$,
which is renormalized by the laser field, the mapping to the Haldane
model must also include a flux $\phi_{ijk}=\arctan\left(\chi_{ijk}/J_{2,ik}\right)$,
and since both of the quantities $\chi_{ijk}$ and $J_{2,ik}$ depend
on $\alpha a_{i}$ in distinct manners. This phase becomes tunable
with the electric field intensity, leading to another mechanism for
tuning the topological phase. In this case, the phase diagram acquires
two interesting features showcased in Fig. \ref{fig:Phase-diagrams-showcasing-1},
especially evident for strain applied in the AC direction. A critical
value of electric field exists which provides a transition for vanishing
values of strain for a much lower value of electric field $\alpha a_{0}\approx1.38$.
The value for the critical electric field is also decreased until
it vanishes, at a strain of about $15\%$. Tuning the strain with
high precision near this value can allow for an inversion of the sign
of the Floquet Chern number for an arbitrarily small electric field.
In the ZZ direction, a similar situation occurs for topological to
trivial transition, near $11\%$ strain, and this value can be reduced
to about $9.5\%$ strain while remaining in the $C_{-}^{F}=-1$ phase
by increasing $\alpha$. {Furthermore, as is clear from the inspection and comparison of the band structure of the FMCI in a non-trivial phase of Figs. \ref{fig:Phase-diagrams-showcasing} and \ref{fig:Phase-diagrams-showcasing-1}, the presence of a $J_2$ term leads to a stabilization of the topological phase. Even for small electric fields, this term ensures a much larger gap will appear, thus rendering the FMCI far more amenable to experimental realizations.} We consider these to be the most important
results of this work, as small values of strain can be achieved by
overlaying a 2D magnetic material in a patterned substrate, and thus
precise control of topological phases can be obtained in integrated
devices.

\begin{figure}
\begin{centering}
\includegraphics[scale=0.4]{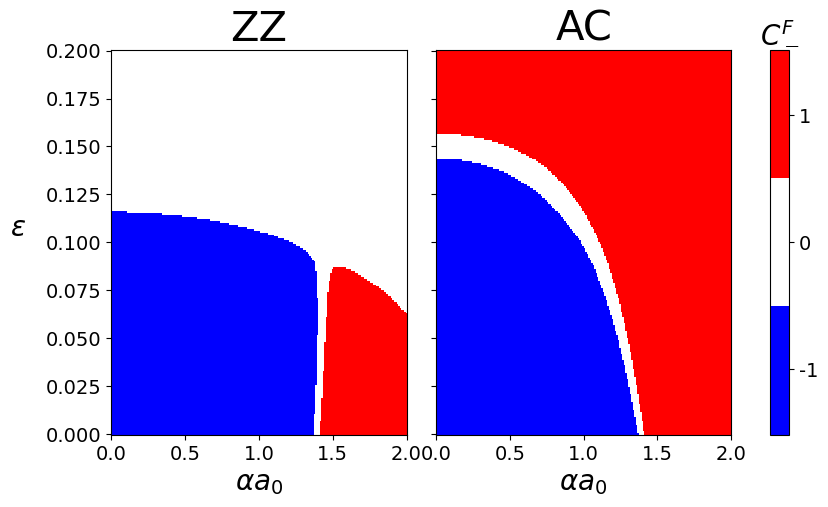}
\par\end{centering}
\caption{Phase diagrams for the Floquet Magnon Chern insulator with a finite
mass $M=3.3\times10^{-3}JS$. For these images we pick {$\beta=3.37$}
and $\nu=0.165$.\label{fig:(a)-Phase-diagram-1}}
\end{figure}

Another aspect which is also worth mentioning is that in the limit
where $M\to0$, the transitions between phases with opposed Chern
number occur directly, as can be seen in Fig. \ref{fig:Phase-diagrams-showcasing-1}
of the main text. On the other hand, for $M>0$, finite regions of
trivial phases crop up in between those characterized by Floquet Chern
number $C_{-}^{F}=-1$ and $C_{-}^{F}=+1$, as seen in Fig. \ref{fig:(a)-Phase-diagram-1}.
This results in an intermediate phase with $C_{-}^{F}=0$, which may
be useful in an experimental context, as it provides a clear barrier
between distinct topological phases. 

\section{Parameter values and realistic realization}

\begin{figure}
\includegraphics[scale=0.4]{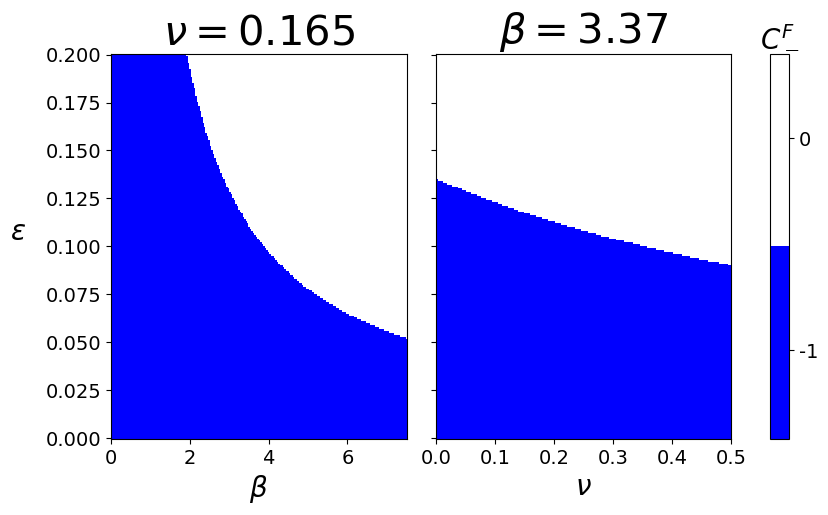}
\caption{{Phase diagrams for the Floquet Magnon Chern insulator with a small applied electric field $\alpha a_0 = 0.01$ and varying strain in the ZZ direction as well as parameters $\beta$ (left) and $\nu$ (right). When one such parameter is varied the other is kept constant at the values considered in the remainder of the text. Increasing either $\beta$ or $\nu$ results in the lowering of the critical strain which induces a topological phase transition.}}\label{fig:Phase-diagrams-for-1}
\end{figure}

We can now review and justify the choice of parameters used for our
calculations, since some are yet undetermined experimentally at the
present time. Parameters such as the Poisson ratio $\nu$ for monolayer
ferromagnetic materials, for instance, have not yet been subject to
thorough experimental analysis, and hence for a rough estimate of
the effects of the strain we have used $\nu=0.165$ corresponding
to the case of graphene, the most well known 2D material with a honeycomb
structure \cite{pereira_tight-binding_2009}. We also pick $\beta=3.37$,
corresponding to the value obtained experimentally for graphene \cite{castro_neto_electron-phonon_2007}.
The electric field at which the first transition occurs for small
strain lies around $\alpha a_{0}\approx2.3$, but is lowered for increased
values of strain in the AC direction in the case of $J_{2}=0$. Using
the distance between magnetic atoms in CrI$_{3}$ as a rough estimate,
a critical electric field of the order of $E_{0}\approx1\times10^{13}\mathrm{V/cm}$
is necessary to induce a transition. Lasers of up to $10^{23}\mathrm{W/cm}^{2}$
have been reported \cite{yoon_realization_2021}, which allow for
laser fields of up to roughly $E_{0}\approx9\times10^{12}\mathrm{V/cm}$.
Although this value is of the order of magnitude of the field necessary
to induce topological phase transitions in the system, the authors
recognize that it still is quite a high value of electric field, which
may result in damage to the material or otherwise undesirable out-of
equilibrium phenomena to take place. This renders the topological
phase transitions in a $J_{2}=0$ model, likely out of reach. However,
as our calculations show, for $J_{2}>0$, the critical electric fields
are much smaller{, and the topological gap is stabilized, facilitating an experimental implementation in essentially every regard.} For increasing values of strain in the AC direction
up to a critical value of $15\%$ a transition occurs for vanishing
field intensity {(see Fig. \ref{fig:Phase-diagrams-showcasing-1})}. In the ZZ direction, a trivial phase can be reached
for values of up to $11\%$ strain. Furthermore, we use $\hbar\omega=50J$,
which leads to a frequency of $\omega/2\pi\approx1.2\times10^{13}$Hz,
lying in the 10s of THz, achievable using ultra-fast terahertz spectroscopy
\cite{owerre_floquet_2017}. {It should also be noted that, for 2D ferromagnetic materials, our choice of parameters is a conservative estimate. Since graphene is known to have very strong carbon-carbon bonds, it is expected that  realistic values of $\beta$ may be much larger for other relevant materials, compatible with a quicker decay of electronic bond strengths. This actually reduces critical strain values. For instance, if $\beta$ is doubled, strain in the ZZ direction can cause a topological phase transition at magnitudes as low as $5\%$. On the other hand, it may be the case that $\nu$ is actually smaller, and this would, in turn, result in an increased critical strain magnitude. This points to a necessity of further exploring elastic properties of 2D magnetic materials. The variation of the critical strain magnitude in the ZZ direction with both $\beta$ and $\nu$ is given in Fig. \ref{fig:Phase-diagrams-for-1}}. {It is also worth noting that increasing the strength of $J_2$ does not alter the phase diagrams presented in any way, although it does increase the magnitude of the topological gap. Hence, if a material actually exhibits a greater value of $J_2$, this can ameliorate the conditions for a physical implementation of SETS.}

Finally, {and in order to discuss an alternative system in which our ideas could be tested,} it is worth mentioning that
the coupling between electrons and electric fields is much stronger.
A number of papers have proposed the realization and study of photo-induced
spin-liquid ground states starting from a Fermi-Hubbard model, realizable
in cold atom lattices or some van der Waals materials \cite{vinas_bostrom_light-induced_2020,bukov_schrieffer-wolff_2016,mentink_ultrafast_2015,claassen_dynamical_2017}.
Such a system would be described by a Hamiltonian of the form

\begin{equation}
H=-t\sum_{\left\langle i,j\right\rangle }e^{i\theta_{ij}(t)}c_{i\sigma}^{\dagger}c_{j\sigma}+U\sum_{i}\hat{n}_{i\uparrow}\hat{n}_{i\downarrow},
\end{equation}
where the time-dependent Peierls phases $\theta_{ij}(t)$ now couple
to electronic creation and annihilation operators $c_{i\sigma}^{\dagger}/c_{i\sigma}$.
Here $U$ is the Hubbard on-site coulomb interaction which is proportional
to the number of electrons $\hat{n}_{i\sigma}=c_{i\sigma}^{\dagger}c_{i\sigma}$
with opposite spins $\sigma=\uparrow,\downarrow$ occupying any given
lattice site. A similar approach to the high-frequency approximation
can be considered, in the spirit of the Schrieffer-Wolff transformation,
where $U$ is treated at the same level as the frequency $\hbar\omega$.
The resulting effective Hamiltonian also exhibits a spin chirality
with similar dependencies on Bessel functions \cite{vinas_bostrom_light-induced_2020}. {Thus, an effective Heisenberg model with topological properties can be obtained, and the manipulation of its topological properties would proceed in exactly the same manner as we have described along this work. The advantage in our model is that the direct coupling to the spin system provides a much simpler and essentially physically equivalent treatment of the topological spin system, with the main difference being that} directly coupling to electrons yields a number of advantages {which may prove relevant} for physical implementations: Firstly,
the coupling factors for magnons $\alpha_{m}$ and for electrons $\alpha_{e}$
are related by $\alpha_{m}/\alpha_{e}=10^{-5}$ for a frequency of
$\hbar\omega=1\mathrm{eV}$, enabling the ability to obtain similar phenomenology for electric fields $10^{5}$ times smaller. Additionally,
the driving frequency can be chosen to be sub-gap, i.e. $\hbar\omega<U$,
as well as off-resonant with $U/n$, where $n$ is an integer. This
means the electronic bands will remain at half-filling when driven
by the laser field, thus avoiding material damage. A HSM plus a spin
chirality term can thus remain a valid description of the model under
driving. Finally, electric fields of $E_{0}\approx1\times10^{7}\mathrm{V/cm}$
can be utilized, which are well within reach of experiments, and allow
for $\alpha a_{0}>1$, reaching most of the relevant parameter space
for our proposal. It is our expectation, that by manipulating the
intensity of laser traps, deformed lattices could be realized in this
setting, providing a possible mechanism to test our ideas in a more
controlled environment.

\section{Conclusions and outlook}

In this work we have analyzed a Floquet Magnon Chern Insulator (FMCI)
consisting of a honeycomb 2D ferromagnet upon which a circularly polarized
laser beam is shewn. The FMCI, by itself, can host topological bands
with a quantized Floquet Chern number synthesized by the laser field.
This synthetic topology is not easily tuned, and hence we propose
a strain engineering approach to increase the ability to manipulate
the topological invariant using the laser field. We argue that this
ability can lead to the development of new spintronics based technologies.
Having studied the case of a nearest-neighbor as well as next-nearest
neighbor exchange interaction within the original ferromagnet, we
show that in the latter case, the topological invariant can become
very sensitive to small amounts of strain for certain values of electric
field intensities, and vice-versa. Using our parameters, strain on
the order of 10\% can be used to make a FMCI undergo topological phase
transitions for very small electric field intensities, or equivalently,
using electric fields on the order of $E_{0}\approx1\times10^{12}\mathrm{V/cm}$,
one can use small amounts of strain to generate topological phase
transitions, which enables the possibility of using, for instance,
patterned substrates for locally manipulating topological invariants,
and generating edge spin current circuits.

{Furthermore, existing devices based on topological magnons, such as magnon diodes, beam-splitters or even Mach-Zender type interferometers \cite{Wang_2018_Topological_Magnonics} could be realized by local variations in strain alone, not relying on changes in magnetization, or creation of holes in the material, but rather on SETS.}

We finally argue that systems based on cold-atom traps can function
as a testing ground for these ideas, since coupling of electric fields
to underlying electronic models of magnetism is much stronger than
to spin systems directly due to the nature of the Aharonov-Casher
effect. Within these models, it is nonetheless possible to generate
topological magnetic terms such as a spin-chirality, and strain could
be implemented in a simple manner by deforming the cold atom lattice. 

\section*{Acknowledgments}

The authors thank António Costa and Joaquin Férnandez-Róssier for
comments on this manuscript. T.V.C.A. acknowledges support by the
Portuguese Foundation for Science and Technology (FCT) in the framework
of the project CERN/FIS-COM/0004/2021 and the hospitality of LIP where
this work was conducted. N.M.R.P. acknowledges support by the Portuguese
Foundation for Science and Technology (FCT) in the framework of the
Strategic Funding UIDB/04650/2020, COMPETE 2020, PORTUGAL 2020, FEDER,
and through projects PTDC/FIS-MAC/2045/2021, EXPL/FIS-MAC/0953/ 2021,
and from the European Commission through the project Graphene Driven
Revolutions in ICT and Beyond (Ref. No. 881603, CORE 3). Additionally, N.M.R.P. 
acknowledges support from the Independent Research Fund Denmark (grant no. 2032-00045B) and  the Danish National Research Foundation (Project No.~DNRF165).

%

\end{document}